\documentclass[11pt,a4paper]{article}
\usepackage[koi8-ru]{inputenc}
\usepackage{natbib}
\usepackage[english]{babel}
\usepackage{amsmath}
\usepackage{graphicx}
\usepackage{indentfirst}
\usepackage{aas_macros}
\newcommand{\md}{\mbox {$\dot{M}$}}
\newcommand{\pyr}{\mbox {${\text{yr}^{-1}}$}}
\newcommand{\myr}{\mbox {~${\rm M_{\odot}\,{\text{yr}^{-1}}}$}}
\def\apgt{\ {\raise-.5ex\hbox{$\buildrel>\over\sim$}}\ }
\def\aplt{\ {\raise-.5ex\hbox{$\buildrel<\over\sim$}}\ }
\newcommand{\ms}{\mbox {$M_\odot$}}
\newcommand{\rs}{\mbox {$R_{\odot}$}}
\newcommand{\ace}{\mbox {$\alpha_{ce}$}}
\newcommand{\at}{\mbox {$\alpha_{th}$~}}
\newcommand{\mc}{\mbox {$M_{Ch}$}}

\def\m{^m\kern-7pt .\kern+3.5pt}
\def\sna{SN\,Ia}

\begin{document}

\begin{center}
{\bf{\Large  Type Ia Supernovae in semi-detached binary systems}}

\bigskip

{\large\textbf A. V. Fedorova, A. V. Tutukov, L. R. Yungelson}

\bigskip
Institute of Astronomy of RAS\\
Pyatnitskaya 48, Moscow 119017, Russia\\

\medskip 
afed@inasan.rssi.ru,
atutukov@inasan.rssi.ru,
lry@inasan.rssi.ru
\end{center}

\noindent
{\bf Abstract.} We have considered  scenarios for the evolution of close binaries resulting in the formation of semi-detached systems in which a white dwarf can accumulate Chandrasekhar mass by accretion from a main-sequence or subgiant companion with $M\sim 2\,\ms$. These white dwarfs, probably, explode as type Ia supernovae or collapse with formation of neutron stars. We have carried out a population synthesis study for these systems and have estimated the occurrence rate of such events in the Galaxy, depending on the parameter of common envelopes, mass-exchange rate in the binary, reaction of the main-sequence component on accretion of helium in the intervening phase of evolution.  We have found that the model occurrence rate of SNe\,Ia in  semi-detached systems is $\simeq 0.2 \times 10^{-3}$\,yr$^{-1}$, i. e., it does not exceed $\sim 10\%$ of the observational estimate of the Galactic occurrence rate of SNe\,Ia. 

\section{Introduction}
\label{sect:intro}
The problem of progenitors of type Ia Supernovae (SNe\,Ia) is not solved as yet.
Relatively high fraction of so-called peculiar SNe\,Ia [possibly, up to $\simeq 40\%$, \cite{li+01}] may suggest that the progenitors of SNe\,Ia may form a non-homogeneous group. Usually, three possibilities are considered: (i) explosion of a white dwarf that has accumulated Chandrasekhar mass \mc\ by accretion in a semi-detached \citep{wi73,it84a} or detached  \citep{ty_symb76,munrenz92} binary system; (ii)  explosion of a merger product of a pair of white dwarfs with (super)Chandrasekhar total mass \citep{ty81,web84,it84a}; (iii) explosion of a (sub)Chandrasekhar mass white dwarf initiated by the detonation of helium in the layer of accreted matter in a semi-detached system with helium donor [edge-lit detonation, \citet{livne90}].

In the present paper we apply population synthesis for close binary stars for the  estimation of possible occurrence rate of SNe\,Ia and accretion induced collapses (AIC) due to accumulation of Chandrasekhar mass by white dwarfs in semi-detached systems with hydrogen-helium donors. We perform a detailed analysis of the dependence of results on the parameters of calculations and show that in the Galaxy $\aplt 10\%$ of all \sna\ may occur in these systems.

\section{Scenarios of evolution to \sna}
\label{sec:scen}

\begin{center}
\begin{figure}[ht!]
\includegraphics[scale=0.6]{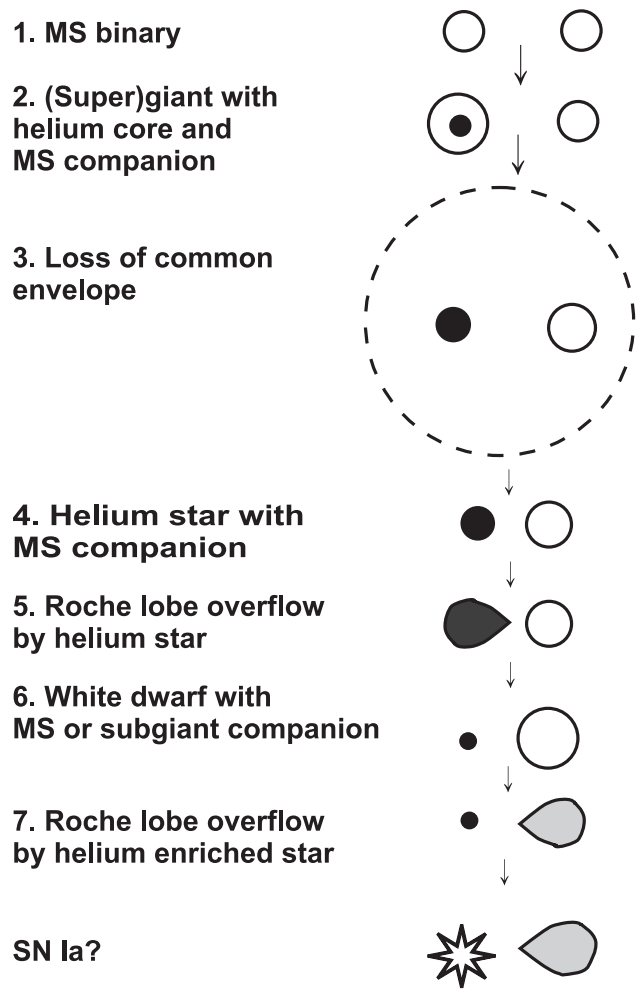}
\caption[]{Schematic representation of a scenario of evolution resulting in the formation of a potential \sna\ in the case B of mass-exchange (``a helium-rich supersoft X-ray source channel'', scenario I).}
\label{fig:fig1}
\end{figure} 
\end{center}

\begin{center}
\begin{figure}[ht!]
\includegraphics[scale=0.6]{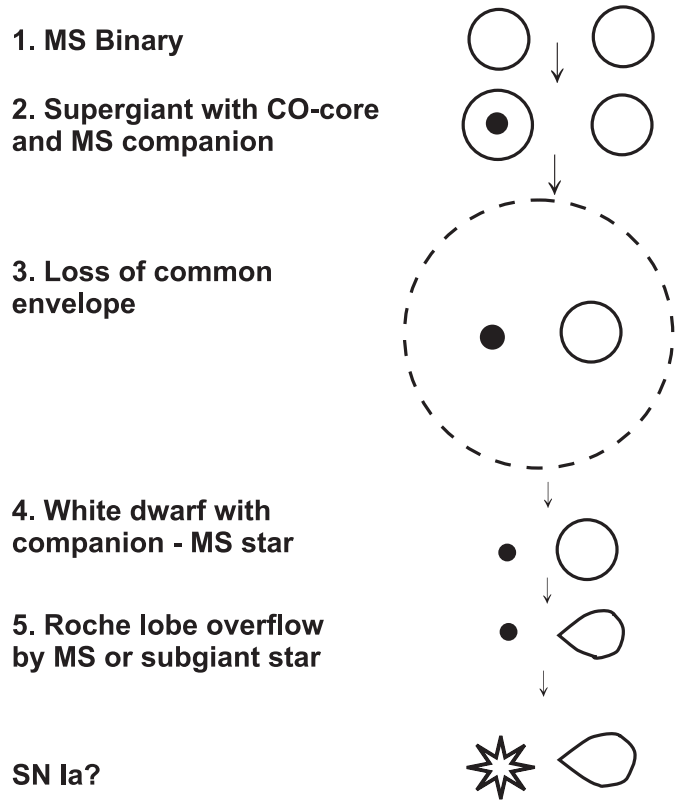}
\caption[]{Schematic representation of a scenario of evolution resulting in the  formation of a potential \sna\ in the case C of mass-exchange, scenario II.}
\label{fig:fig2}
\end{figure}
\end{center}

Scenarios of evolution of close binaries to \sna\ are sketched (out of scale) in Figs. \ref{fig:fig1} and \ref{fig:fig2} (hereafter, scenarios I and II, respectively). In both scenarios initial mass of the primary component is $M_{10} \approx 5 - 11$\,\ms, mass of the secondary is $M_{20} \approx  0.8 - 3.5$\,\ms. In scenario I separation of components corresponds to the RLOF by the primary in the hydrogen-shell burning stage (case B of mass exchange). Mass loss by the primary results in formation of a common envelope. Evolution inside common envelope leads to the shrinkage of the orbit. Due to the mass loss, primary component becomes a helium star with a mass $\simeq (0.8 - 2.5)\,\ms.$ Helium stars of this mass range expand after formation of the CO-core \citep{it85,bitzaraki02}. This results in the second RLOF and accretion of helium-enriched matter by the companion. Mass loss by an expanding star with a CO-core and a helium envelope occurs in the thermal time scale of the star ($\sim\!10^5\,\text{yr}$). Applying results of evolutionary computations by \citet{it85} one may estimate the mass of the lost envelope
\begin{equation}
\label{eq:deltam}
\frac {\Delta M} {M_\odot} \approx 0.2 \left ( \frac {M_{10}} {7 M_\odot} \right )^4
\end{equation}
and mass-loss rate
\begin{equation}
\label{eq:mdothe}
\md_{He} \approx 6.3 \times 10^{-13} \left ( \frac {M_{10}}{M_\odot} \right )^{7.5} \,\myr.
\end{equation} 
As a result of mass loss, helium star becomes a CO white dwarf. Accretion, most probably, leads to the complete mixing of the secondary (see below). The latter becomes somewhat helium-enriched:  $Y\sim 0.4$. [\citet{hknu99} dubbed this scenario ``a helium-rich supersoft X-ray source channel'']. The secondary fills its Roche lobe in the main-sequence stage or soon after developing a helium core. If ensuing mass transfer is stable, accreting dwarf may accumulate Chandrasekhar mass.

Scenario II differs from scenario I by larger initial separation of components. Consequently,
the primary component fills its Roche lobe in the AGB and becomes a white dwarf avoiding the stage of helium star. 

In both scenarios in the phase of accretion onto a white dwarf the system may manifest itself as a source of supersoft X-rays \citep{kylafis_lamb82,hbnr92}.  

In scenario I one of the stages of the evolution is the phase of accretion of helium from the envelope of the primary remnant (Fig.~\ref{fig:fig1}, stage 5). In our previous papers we have assumed that, due to high accretion rate that is defined by Eq.~(\ref{eq:mdothe}), a common envelope that is completely lost by the system  is formed in this phase. On the other hand, for instance, \citet{hknu99} assumed that common envelope does not form and mass exchange is completely conservative. Actually, the possibility of formation of common envelope depends on the reaction of the radius of accreting star on mass gain. Thus, this reaction becomes a parameter of the problem. It becomes necessary to study the dependence of the model rate of \sna\ upon behavior of the accretor radius.  

\section{Accretion of helium by a main-sequence star with initial mass of 2\ms}
\label{sec:accr}

The possibility of accretion of helium, i. e., of the matter with molecular weight $\mu$ exceeding that at the surface of the main-sequence accretor, was noticed long ago \citep{stot_simon69}. However, evolution of low- and moderate-mass stars accreting He was not studied as yet.

We have considered the evolution of a main-sequence star with initial mass $M_0=2$\,\ms\ that accretes $\Delta M = 0.5$\,\ms\ of helium-rich matter with a constant rate $\md_a$. This combination of $M_0$ and $\Delta M$ corresponds to a quite typical close binary system with initial mass of components 9 and 2 \ms\ that may be a progenitor of a \sna\ [see Eq. \eqref{eq:deltam} and  Fig.~\ref{fig:fig5} below].

Computations were carried out for several values of $\md_a$ (Table~\ref{tab:tracks}). The choice for the set of $\md_a$ was defined by the possibility of realization of these rates in the evolutionary scenarios resulting in \sna\ [see Eq.~\eqref{eq:mdothe} and Fig.~\ref{fig:fig5}]. 
We used for the computations the evolutionary code designed for the studies of low-mass stars \citep{tfeng01}. Initial chemical composition of accretor was $X = 0.70, Y=0.28, Z=0.02$, the chemical composition of accreted matter -- $Y=0.98, Z=0.02$. 

We have considered two sets of assumptions on the evolution of accreting star.

An inverse gradient of molecular weight has to lead to the instability and mixing at the border of helium layer \citep[see, e. g.,][]{ulrich72}. \citet{kipp_thomas_80} have shown that for a non-rotating spherically-symmetric star the mixing may be considered as formation of helium ``bubbles'' and their diffusion inside hydrogen-rich substratum. Sinking bubbles are gradually destroyed and mixed with surrounding matter. In this model, diffusion coefficient $D$ is 
\begin{equation}
\label{eq:kipp}
D = \frac {H_p} {\nabla_{ad} - \nabla} \frac {4acT^3} {C_p\kappa \rho^2}
\Bigl| \frac{d\mu}{dr} \Bigr| \frac {1}{\bar{\mu}}.
\end{equation}
Here $\kappa$ is opacity,
$\bar{\mu}$ is the average molecular weight in the region with the gradient of molecular weight and we use the standard notation for the rest of physical variables. The time scale $\tau$ for the doubling of the thickness $W$ of the transitional layer with the gradient of molecular weight is \citep{kipp_thomas_80}
\begin{equation}
\label{eq:tau}
\tau \approx W^2/D.
\end{equation}
In the first set of computations we have used expressions (\ref{eq:kipp}) and (\ref{eq:tau}) for determination of the variation of the distribution of He in the star in the course of accretion.

In the second set of computations we have assumed that the mixing of accreted matter occurs much faster than it was found by \citet{kipp_thomas_80}. This assumption was based on the estimates of the time scales of physical processes that accompany the sinking of turbulent eddies. These time scales, at least in 1D approximation, are much shorter than the time scale of accretion $\sim (10^5 - 10^6)$\,yr (see Appendix). Therefore, we have carried out a set of computations under assumption of an ``instantaneous'' mixing of accreted matter.

\begin{figure}[ht!]
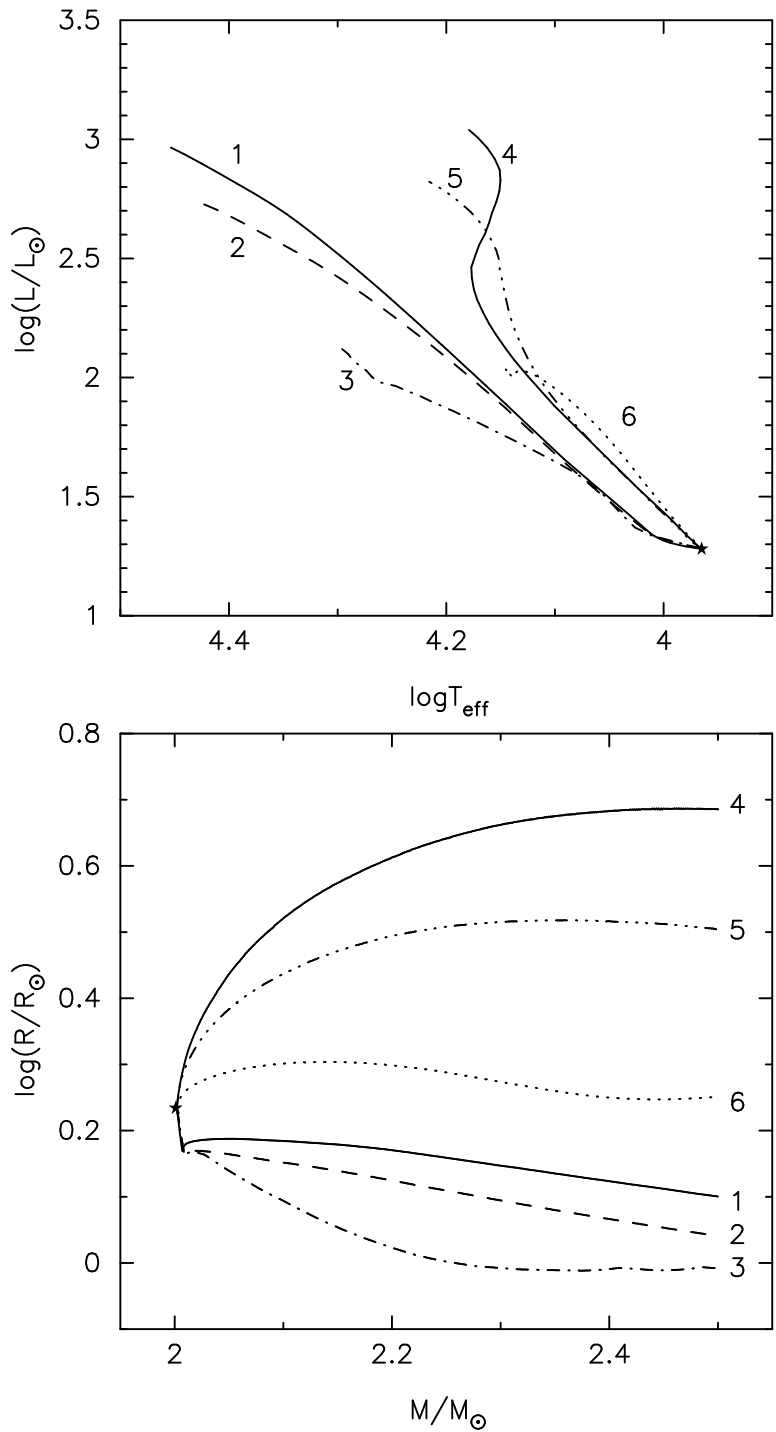

\vspace{-2cm}
\includegraphics[scale=0.5,angle=-90]{hreng.ps}
\includegraphics[scale=0.5,angle=-90]{ris_R_neweng.ps}
\caption[]{Variation of parameters of accreting stars depending on accretion rate and mode of mixing. Upper panel -- tracks of the stars in the HR diagram. Lower panel -- dependence of radii of accreting stars on mass. Notation corresponds to the numbers of tracks in Table~\ref{tab:tracks}. The  position of initial model is marked by an asterisk.}
\label{fig:fig3}
\end{figure}

\begin{figure}[ht!]
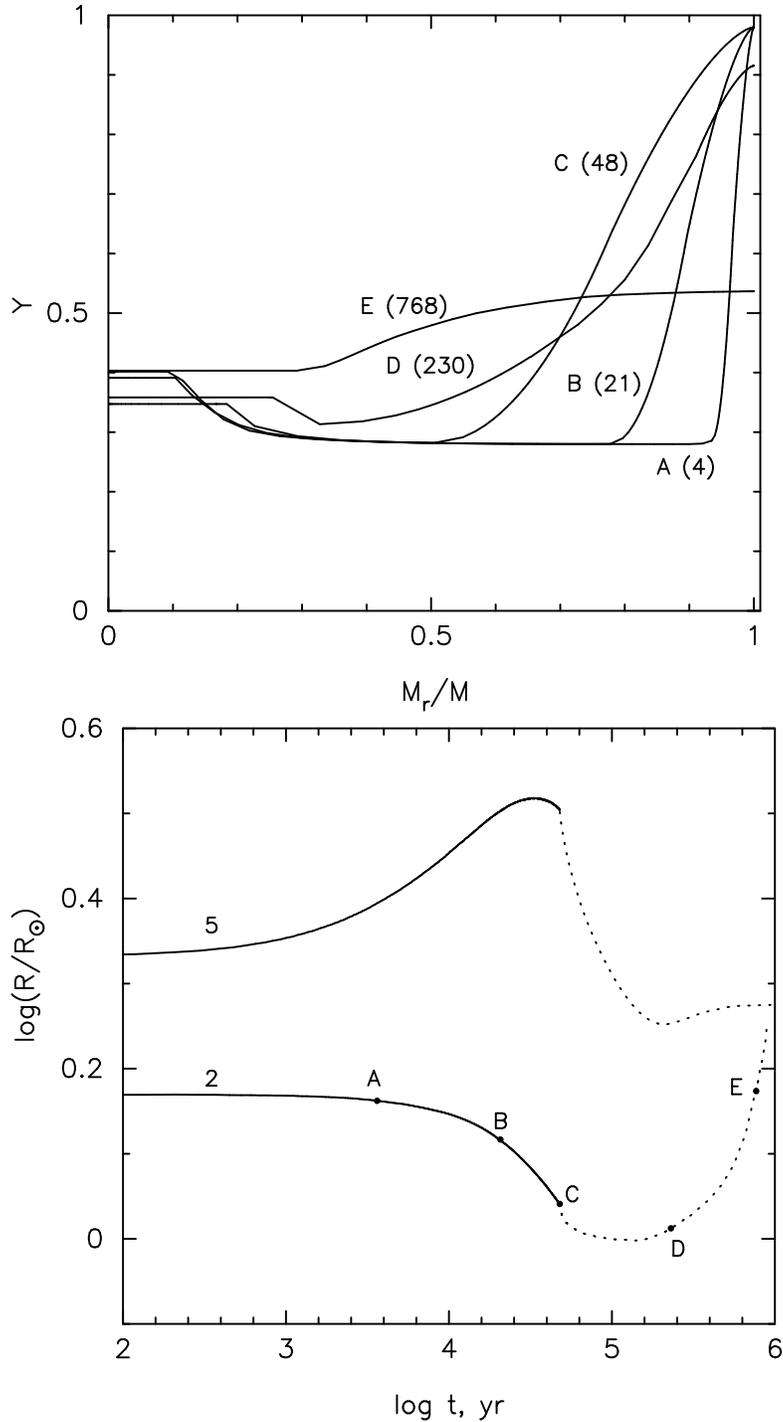

\includegraphics[scale=0.5,angle=-90]{profeng.ps}

\includegraphics[scale=0.5,angle=-90]{rteng.ps}
\caption[]{
Upper panel: variation  of the profile of helium concentration in the models of track 2 with time (including the stages after termination of accretion). The numbers indicate time that elapsed since beginning of accretion in $10^3$\,yr. Lower panel: variation of stellar radii in the models of tracks 2 and 5 (solid curve corresponds to the stage of accretion while the dotted curve -- to the evolutionary stage after termination of accretion. The letters indicate the models with helium profiles shown in the upper panel.  }
\label{fig:fig4}
\end{figure}

The parameters of evolutionary tracks are listed in Table~\ref{tab:tracks}. The tracks of accreting stars in the Hertzshprung-Russell diagram and variation of their radii are shown in Fig.~\ref{fig:fig3}.  In the upper panel of Fig.~\ref{fig:fig4} we show (for track no. 2) the variation of the profile of helium distribution inside accreting star with time; this plot also partially includes
the phase of evolution after cessation of accretion. Computations have shown that under assumption of diffusion (tracks 1 -- 3) the radii of stars decrease in the course of accretion due to the increase of transparency of the matter with accumulation of helium.

An expansion of the star that is usually typical for main-sequence stars that  accrete matter at a high rate occurs under assumption of ``instantaneous'' mixing
(tracks 4 -- 6). For accretion rate $2 \times 10^{-5}\,\myr$, the maximum radius of the star is almost three times larger than the initial one; this ratio is lower for lower accretion rates (Table~~\ref{tab:tracks}). The increase of stellar radii is due to the liberation of internal energy with increase of the mean molecular weight of their matter. 

After termination of accretion phase, all models irrespective of mixing mode relax to almost similar radii in the thermal time scale (Fig.~\ref{fig:fig4}, lower panel).

Thus, evolutionary computations still do not allow to decide definitely what is the character of the variation of radii of  1 -- 3\,\ms\ stars that accrete helium at evolutionary defined rates. However, it is clear that the assumption about formation of common envelopes due to accretion of helium that was applied in our earlier studies is not the only option.

\section{Population synthesis for progenitors of \sna\ in semi-detached  binaries}
\label{sec:synth}

In our previous papers we have studied the occurrence rate of accumulation of \mc\ by white dwarfs in the systems of different types: cataclysmic variables that are evolving in the thermal time scale of the donor or in the time scale of angular momentum loss via magnetically coupled stellar wind \citep{ylttf96}; in the systems with subgiant donors with degenerate helium cores that evolve in the thermal or nuclear time scale of the donor \citep{ylttf96,yl98,yl00}. In the former systems mass of the donors was limited from above by 1.5\,\ms, while in the latter  by 2.5\,\ms.  Our estimates of the occurrence rate of potential \sna\ in semi-detached systems did not exceed $\sim 10\%$ of the empirical estimate of the \sna\ rate in the Galaxy that is $(4\pm2)\times10^{-3}\,\pyr$\ \citep{captur01}.
On the other hand, \citet{hkn96,hknu99} and  \citet{lh97} have found, that the model rate for the systems of the same type may be comparable to the empirical one. We should note, however, that the starting point for the estimates made in the papers of Hachisu et al. and Li and van den Heuvel were the systems composed by white dwarfs with main-sequence or subgiant companions. The authors of the aforementioned  papers did not estimate the rate of formation of these ``progenitor systems'' by means of population synthesis. This may be one of the causes for the discrepancy \citep{yl98}.

In the present paper we study the possibility of accumulation of \mc\ by white dwarfs evolving according to scenarios I and II, using uniform assumptions on the mass loss by the donor and the possibilities of mass gain by white dwarfs.   It is evident that most part of the matter potential SNe\,Ia accrete in the thermal time scale of the donor, when the hydrogen burns stationary and the rate of its accumulation is $\md_{st} \simeq 10^{-7} - 10^{-6}\,\myr$. In addition to binary systems considered by us before, we have included into consideration binaries with main-sequence components that satisfy the condition of dynamically stable mass exchange: $q = M_2 / M_1 \aplt 2.5$ \citep{tfy82,hw87}. If $1.2 \aplt q \aplt 2.5$, mass exchange in these systems proceeds in the time scale close to the thermal time scale of the donor until masses of components become equal.   
Accretion rate $\md_a$ may be then higher than the maximum rate of stationary hydrogen burning $\md_{st}$. Following \citet{katoiben92,kathac94,hkn96}, we have assumed that the stationary burning of the hydrogen generates optically thick stellar wind that takes excess of the matter away from the system. Specific angular momentum of the wind matter is equal to the specific angular momentum of the dwarf. Mass loss in this mode stabilizes mass exchange in the system. Optically thick wind exists only if  $\md_a \aplt 10^{-4}$\,\myr. For higher $\md_a$ common envelopes should form.

\begin{center}
\begin{figure}
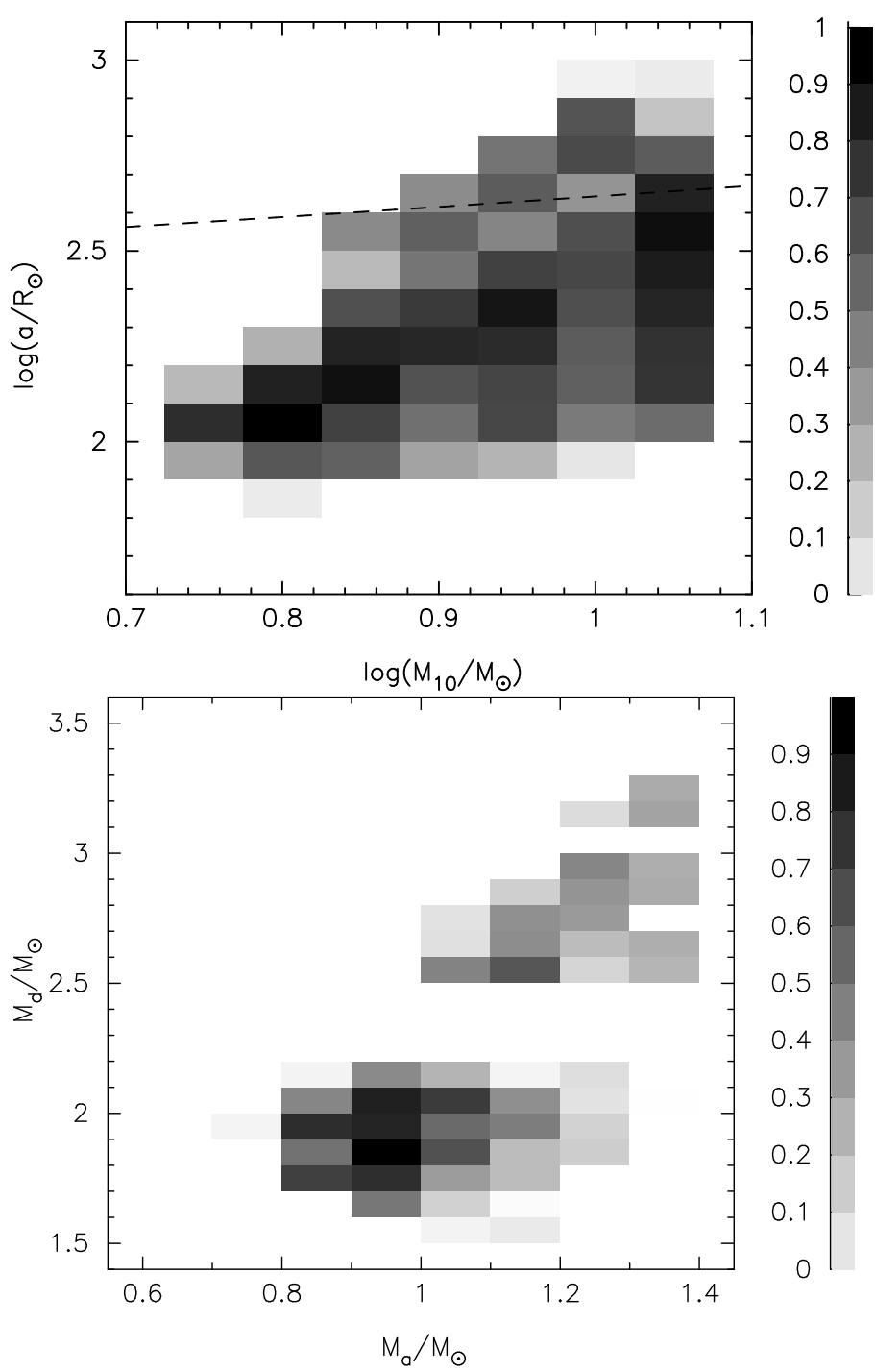

\vspace{-2.5cm}
\includegraphics[scale=0.5,angle=-90]{ma_aeng.ps}
\vspace{1cm}
\includegraphics[scale=0.5,angle=-90]{ma_mdeng.ps}
\caption[]{Upper panel: progenitors of systems in which white dwarf can accumulate \mc\ in the ``initial mass of the primary -- initial separation of components'' diagram. The border between systems experiencing cases B and C of mass exchange (for $M_{20}=2\,\ms$) is shown by dashed line. The maximum of gray scale corresponds to the systems that provide \sna\ occurrence rate of  $0.94\times 10^{-5}$\,yr$^{-1}$. Lower panel: the relation between mass of the donors and accretors in the systems in which white dwarf attains \mc\ at the moment when the donor overfills its Roche lobe (beginning of the stages 7 and 5 in scenarios I and II, respectively). The maximum of gray scale corresponds to the systems that provide \sna\ occurrence rate of  $0.15 \times 10^{-4}$\,yr$^{-1}$. The figure corresponds to model 13 in Table~\ref{tab:scen}.
 }
\label{fig:fig5}
\end{figure}
\end{center}

To improve the accuracy of the computations of the potential occurrence rate of \sna\ in semi-detached systems, we have introduced some modifications concerning efficiency of accumulation of the matter by white dwarfs, mass loss from the systems, and behavior of the radii of the accreting stars into the population synthesis code we have used before  \citep{ylttf96,yl98,yl00}. Here we mention only that we assume that all stars in the Galaxy are born in binaries and that we apply the birthrate function of binary stars in the form suggested by \citet{it84a} 
\begin{equation}  
\label{eq:brate}
\frac{dN}{dt} = 0.2 d (\log a_0) \frac {dM_{10}}{M_{10}^{2.5}} f(q_0) dq_0,   
\end{equation}
where $a_0$ is the initial separation of components, $f(q_0)$ --  distribution of initial mass ratios of components normalized to 1 ($q_0=M_{20}/M_{10}$).
Function \eqref{eq:brate} is normalized to the formation of one binary system with 
$M_{10} \geq 0.8\,\ms$ per yr. The mass of progenitors of white dwarfs is (0.8 -- 11.4)\ms\ \citep{it85}.
It was assumed that star formation rate in the Galactic disk was constant for $13\times10^{9}$ yr\footnote{This assumption is justified since we consider rather short-living ($\sim 10^9$ yr) stars with $M\apgt1.6$\,\ms\ (see Fig~\ref{fig:fig5}).}. Initial distributions of binaries over $q_0$ and $\log a_0$ were assumed to be flat. The space of initial parameters of the binaries was approximated by rectangular grid with steps 0.0125 in $\log M_{10}$,  0.0125 in $\log(a_0)$, and  0.025 in $q_0$.

The number of systems with mass exchange resulting in the accumulation of \mc\ depends, mainly, on the parameter of the common envelopes \ace. The ``equation of common envelopes'' was taken in the form 
\begin{equation}
\frac{ (M_{a0} + M_{d0}) \Delta M}{a_0}=\ace \left[ \frac{M_{df}M_{af}}{a_f}-\frac{M_{a0} M_{d0}}{a_0} \right],
\label{eq:ce}
\end{equation} 
where indices  $0$ and $f$  indicate initial and final values of the masses of donor and accretor and their separations and $\Delta M$ is the mass of the lost envelope. The value of \ace\ is the measure of the ratio of the binding energy of the common envelope and the orbital energy of the binary system. At present, neither theory nor observations allow to restrain \ace\ with reasonable accuracy. Our choice of the range of \ace\ used in the computations (0.5 to 5) is justified by the results of the analyses of observations \citep{iben2000,nvy+00,soker_ce02} that provide evidence for a high efficiency of the deposition of energy into common envelopes and for the loss of a considerable part of the donor envelope by stellar wind prior to RLOF, that is equivalent to high \ace. Of course, it should be noted that the usage of the same \ace\ for all common envelopes is a forced simplification of the problem [see \citet{dt00,td01} where the binding energy of the envelopes is evaluated as the function of the evolutionary state of the star].

For every semi-detached system  with main-sequence donor that was generated by population synthesis code, mass loss rate was assumed to be constant and equal to $\md_d = \alpha_{th}\dot{M}_{\mathrm {KH}}$, where $\dot{M}_{\mathrm {KH}}$ is the mass loss rate corresponding to the thermal time scale of the donor.  It is evident that in real systems $\md_d$ depends on the mass ratio of components (decreases with decreasing mass ratio of components) and on the evolutionary state of the donor at the instant of the RLOF. As well, it varies in the course of evolution. The value of \at\ is, thus, an additional parameter of the model . The value of \at\ was varied from 0.2 to 1. The computations were continued until white dwarf accumulated \mc\ or while mass of the donor became equal to the mass of accretor. In the systems with subgiant donors mass loss first occurred in the thermal time scale of the donor and then -- in the time scale determined by the rate of growth of the donor helium core \citep[see, e. g.][]{yl98}.  In these systems computations were limited either by accumulation of \mc\ by the white dwarf or by complete loss of the hydrogen envelope by the donor.

For the estimate of the efficiency of helium accumulation by white dwarfs (or their erosion due to Novae explosions at $\md_a \aplt 1.5\times 10^{-8}\myr$), the grid of models of \citet{prkov95} was applied. The loss of matter via thermal flashes in the helium surface layer of white dwarf accumulated due to hydrogen burning was estimated according  to \citet{kh99}.

According to the results of computations of accretion (Sec.~\ref{sec:accr}) we have considered two limiting cases of reaction of the main-sequence star -- future donor in the system with white dwarf -- on accretion of helium-enriched matter:
completely conservative mass exchange with the radius varying according to the mass-radius relation only, and conservative mass exchange that is accompanied by a three-fold increase in accretor radius. For the sake of comparison with our previous studies we also considered the occurrence rate of \sna\ assuming that accretion is accompanied by formation of a common envelope.

Results of computations are given in Table~\ref{tab:scen}.
 For every combination of \ace\ and \at\ we give separately the occurrence rate of possible \sna\ accumulating \mc\ according to scenario I under three different assumptions on the reaction of the star upon accretion (three upper lines). The fourth line for every set of parameters in the Table gives the rate of events in the systems evolving through scenario II. Separately, as AIC, are listed accumulations of \mc\ by white dwarfs with initial mass $\apgt 1.19$\,\ms\ that, under our assumptions, consist of the oxygen-neon-magnesium mixture. At accretion rates exceeding $\simeq 10^{-8}$\,\myr\ these dwarfs, perhaps, do not explode as \sna\ but, instead, collapse with formation of neutron stars \citep{nomkondo91}.
``Initial mass of the primary component -- initial separation of components'' diagram for progenitors of the systems in which white dwarfs may accumulate \mc\ is shown in Fig.~\ref{fig:fig5} [for one of the most ``fertile'' model no. 13 in Table~\ref{tab:scen}]. The same figure shows the relation between masses of the donors -- main-sequence stars or subgiants and masses of white dwarfs in the systems in which white dwarfs are able to accumulate \mc,  at the moment when the  mass transfer starts. Two groups of objects in the lower panel of Fig.~\ref{fig:fig5}   are systems with main-sequence donors ($M_d \geq 2.5$\,\ms) and systems with subgiant donors  ($M_d = 1.5 - 2.2$\,\ms). The cause for the  ``gap''
between two groups of objects is formation of deep convective envelopes in the stars with $M \apgt 2.2\,\ms$ immediately after TAMS; this results in dynamical mass loss upon RLOF and formation of common envelopes.

Interestingly, a significant fraction of ``presupernovae'' white dwarfs is initially not very massive  ($M_a \aplt 1\,\ms$) and have relatively low-mass companions ($M_d \aplt 2\,\ms$). The circumstance that in these systems the combination of the mass accretion rate and the stationary burning rate is favorable for accumulation of \mc\ is in agreement with results of \citet{langer+00} and \citet{ergma_usco02} who have carried out detailed computations for the main-sequence star -- white dwarf systems. 

The dependence of results on the parameters of the population synthesis may be summarized as follows.

\begin{itemize}
\item[ ]
The number of the systems with white dwarfs successfully accumulating \mc\ depends, mainly, on \ace. For $\ace > 1$ the number of systems evolving according to scenario I diminishes and for \ace = 2 they practically disappear, because efficient deposition of energy into  common envelopes limits the separation of components from below. As a rule, scenario I is dominating: see Fig.~\ref{fig:fig5} in which the border between the systems evolving according to scenarios I and II is shown; this line only weakly depends on $M_{20}$.

\item[ ]
Reduction of  \at\ from 1 to 0.5 results in increase of the number of potential \sna. For high \at\ a significant fraction of the matter is accreted with the rate that is sufficiently higher than the rate of the stationary hydrogen burning and is, therefore, lost from the system via optically thick wind. For this reason, the efficiency of mass accumulation is low.  For low \at = 0.2, efficiency of helium accumulation is also low, due to the recurrent explosions of Novae; this results in reduction of the number of systems with dwarfs that are able to accumulate \mc.

\item[ ]
In scenario I, for $\ace \ge 1$ and \at $\ge 0.5$ the assumption about formation of a common envelope in the course of accretion of the He-rich matter increases the number of potential \sna, since this increases the number of systems in which RLOF becomes possible. For \ace $ < 1$, on the contrary, the mergers of components reduce the number of potential \sna.

\item[ ]
The effect of a three-fold increase of the radius of the accreting star as compared to the case of mass exchange without expansion of accretor, is comparable to the case with formation of the common envelopes. The reason for the similarity is in the narrow interval of the major semi-axes of orbits in progenitor systems: $\Delta \log\,a_0 \approx 0.3 -  1$ (Fig.~\ref{fig:fig5}).

\item[ ]
Comparison with results of \citet{yl98} shows that taking into account systems with initial mass of the secondary higher than 2.5\,\ms\ results in increase of the \sna\ rate by less than $\sim 30\%$. The increase of mass of the donors more strongly influences the rate of AIC's that almost doubles [c.f. Table~1 in   \cite{yl98}].      

\end{itemize}

Semi-detached systems with accreting white dwarfs that burn hydrogen stationary are usually identified with one of the subpopulations of supersoft X-ray sources \citep{hbnr92,rdss94,ylttf96}. In Table~\ref{tab:scen} for every set of computations we give the number of dwarfs $N_{\rm SSS}$ that may be observed as sources of supersoft X-rays (for scenarios I and II together). For the lifetime of stationary sources we have assumed the time of accretion with $\md \geq \md_{st}$;
for nonstationary sources the lifetime was assumed equal to the period of time when, after Nova explosion, the luminosity of the dwarf exceeds  $4 \times 10^{37}$\,erg\,s$^{-1}$ [see. Eq.~(12) in \citet{ylttf96}]. 

In scenario I that is actual for most systems, the matter of the donors has to be enriched in He ($Y \sim 0.4$). Note, helium-enriched stationary sources of supersoft X-ray emission are not known as yet (P. Kahabka, private communication). The only observed source enriched in He is recurrent Nova U Sco.
The presence of a white dwarf with the mass close to \mc\ is suspected in this system \citep{hac+00,thoro+01}. The estimates of mass exchange rate in this system show that explosions at the surface of the dwarf in this system do not result in erosion and the dwarf may really accumulate \mc\ \citep{ergma_usco02}.

In the Galaxy, only two of the sources of supersoft X-rays that are  identified optically have orbital periods of $\sim 16$\,hr and $\sim 4$\,day and may belong to the subtype considered in the present paper \citep{kahabka02}. The number of sources that may be identified very strongly depends on the ill known interstellar absorption and the distribution of sources over the temperature. \citet{stefrap94b,stefrap95,stefkong03} have shown that in the Milky Way and external galaxies  observed and real number of stationary sources may differ by 2 to 3 orders of magnitude. Given this uncertainty, our estimate of $N_{\rm SSS}$
(Table~ \ref{tab:scen}) does not contradict observations.

\section{Discussion}
\label{sec:disc}

Occurrence  rate of accumulations of \mc\ by white dwarfs in semi-detached systems in our models does not exceed   
 $0.22 \times 10^{-3}$\,yr$^{-1}$ (Table.~\ref{tab:scen}, model 22). Even if we assume that the systems which in our model experience accretion-induced collapses actually explode as \sna, the total rate of events does not exceed $\simeq 0.4 \times 10^{-3}$\,yr$^{-1}$ (model 26); the latter model requires extremely high \ace\ and a kind of ``optimization'' on \at! This model occurrence rate comprises only 20\% of the lower limit of the observational estimate of the occurrence rate of \sna\ in our Galaxy. The model estimate is even somewhat overestimated (by 10 -- 20\%) since, even if the mass transfer rate is within the limits that correspond to the stationary burning of the hydrogen, some of the accreted matter has to be lost via stellar wind from a hot white dwarf. Nevertheless, the fraction of potential \sna\ that can occur in semi-detached systems is not negligible. A critical observational evidence for explosions of \sna\ in low-mass semi-detached systems would be provided by discovery of hydrogen in the spectra of \sna. The main source of the latter should be the matter stripped from the companion. Computations of \citet{marietta00} have shown that for the systems under consideration the mass of stripped matter may be $\sim 0.15$ \ms. Hydrogen may be observed both in very early and very late optical spectra, in radio, and in X-rays  \citep{eck+95,marietta00,lentz+02}. 
An unambiguous evidence in the favor of single degenerate scenario would be provided by discovery of the companion of exploding dwarf. This companion, due to the interaction with the envelope of SN, may have anomalously high luminosity for the first $\sim 10^3 - 10^4$ yr after explosion \citep{marietta00,pods03}. Spatial velocity of the companion after explosion may be up to $\simeq 450$ km/s \citep{canal+01}.

For the estimate of the real role of the channel of semi-detached systems a crucial role may be played by an evaluation of the real number of the supersoft X-ray sources in the Milky Way and external galaxies.

Recently discovered \sna\  2002ic \citep{hamuy03} with hydrogen lines in the spectra is, most probably associated with a massive AGB object and may represent either an explosion of a single star (so called SN\,1.5) or an explosion of a degenerate component in a symbiotic binary system \citep{hamuy03,cy03}.

Our estimates of the occurrence rate of accumulations of \mc\ by degenerate dwarfs are several times lower than the estimates by  \citet{hkn96,lh97,hknu99,hp03} who have found that the model estimates may be more comparable to the observational estimates. Hachisu et al. and Li and van den Heuvel have also used Eq.~\eqref{eq:brate} for the estimate of the formation rate of binary systems, but did not carry out a detailed population synthesis study. Because of nonlinear variation of masses of components and semi-major axes of orbits in the course of evolution, simple replacement of numerical integration of the function \eqref{eq:brate} over the space of initial parameters by multiplication of the finite intervals of $M_{10}$, $\log a_0$, and $q_0$ results in a significant overestimation of the occurrence rate of potential \sna. Another reason for the difference with the results of other studies may be still uncertain efficiency of the matter accumulation by accreting dwarfs.

For the ``standard'' assumption of \ace=1 the same population synthesis code gives $2.1 \times 10^{-3}$\,yr$^{-1}$ for the rate of mergers of close double degenerates with total mass exceeding \mc. For the rate of edge-lit detonations the rate is $4.6 \times 10^{-3}$\,yr$^{-1}$. These estimates are comparable to the observational estimate of the occurrence rate of \sna\ in our Galaxy.

The age of semi-detached systems considered in the present paper  at the moment of \sna\ explosion does not exceed $\simeq 2.5$\,Gyr \citep[see for discussion][]{yl98}. The age of SN initiated by edge-lit helium detonations does not exceed $\simeq 1$\,Gyr \citep{yl00,rego+03}\footnote{It is possible that the lifting effect due to rotation of the dwarf may prevent helium detonation \citep{langer+03}.}. Hence, semi-detached systems cannot produce \sna\ in old populations (elliptical galaxies). The merger of double-degenerates is apparently the only mechanism that is able to produce (super)Chandrasekhar mass objects in the populations of any age. 
 
The existence of close pairs of white dwarfs with total mass close to the Chandrasekhar one and orbital periods short enough for merger in Hubble time  is out of question \citep{nap+02,nap+03,nap_spy03}. The relative number of such pairs is close to the theoretically expected one \citep{nyp+01}: 3 out of $\sim 1000$ field white dwarfs with $V\leq16\m5$ studied for binarity. However, it is still not clear whether  the merger results in the carbon ignition in the center of the dwarf and \sna.

In the merger process, the less massive object turns into a disk that surrounds the ``core''. Further evolution of the system is determined by the rate of accretion from the disk $\md_a$.  The rate of accretion depends on the time scale of angular momentum transfer in the disk $T_a$.   The latter may be of the order of $10^9$\,yr, if it is determined by the viscosity of degenerate electrons only. However, due to the turbulence in the transition layer between the disk and the core, $T_a$ is, most probably, by several orders of magnitude shorter \citep{mochko90} and is comparable to the limiting Eddington accretion rate ($\md_{Edd} \approx 2\times 10^{-5}\,\myr$  for a 1\,\ms\ white dwarf).

The results of accretion computations are ambiguous even in 1D case. \citet{nomoto_iben85,kawai_etal87} have shown that the central ignition of carbon and \sna\ after accumulation of \mc\ is possible for $\md_a\aplt(0.1 -0.2) \md_{Edd}$ only.   For larger $\md_a$, carbon ignites in the layer of accumulated matter and propagates inward. Ignition of carbon initiates the ``competition'' of two processes: propagation of carbon burning and matter accretion from the disk. This allows to get some simple estimates. The maximum rate of the carbon burning, (with account for the fraction   $\alpha_{\nu}$ of released energy that is taken away by neutrinos) is limited by the Eddington luminosity. Accretion rate also does not exceed the limit set by $L_{Edd}$. The dwarf is able to accumulate \mc\ before the burning propagates to the center, if \begin{equation}
\label{eq:mame}
\frac {\mc - M_{a0}} {\mc} \frac {\epsilon_g} {(1- \alpha_{\nu}) X_C \epsilon_C} < 1.
\end{equation} 
Here $M_{a0}$  is the initial mass of accretor, $X_c \approx 0.5$ is carbon abundance in the outer layers typical for massive dwarfs, $\epsilon_C \approx 5 \times 10^{17}$ erg\,g$^{-1}$ is the energy released by burning of 1\,g of carbon, and $\epsilon_g \approx 5 \times 10^{16}$ erg\,g$^{-1}$ is the energy released by accretion of 1\,g of the matter from the disk. Since in the carbon burning layer   $\alpha_{\nu} \sim 0.4$ \citep{sn98}, the estimate \eqref{eq:mame} shows that a dwarf with $M_{a0} = 1\,\ms$\ is able to accumulate  \mc\ before carbon burns completely. Thus, it is possible that ignition of carbon in the outer layers of the dwarf does not prevent central explosion and \sna. Only for $\alpha_{\nu} \apgt 0.9$ complete exhaustion of carbon results in formation of an oxygen-neon-magnesium  white dwarf that, after accumulation of \mc, ``quietly'', without SN, collapses into neutron star.

On the other hand, \citet{dbi03} have shown that, for the accretion rate
 of $5\times10^{-7}$\,\myr\ and  initial mass of CO white dwarfs  (1.33 -- 1.34)\,\ms, carbon burning that starts at the edge of the dwarf and occurs in the regime of delayed detonation is possible.  A \sna\ may result in this case.

 The situation may change if rotation is taken into account. \citet{piersanti+03} have shown that, for certain combinations of accretion rate and spin velocity of the white dwarf, rapidly rotating white dwarf deforms by adopting an elliptical shape. The anisotropic mass distribution induces the loss of rotational energy and angular momentum via gravitational wave
radiation. The white dwarf contracts and achieves the conditions suitable for
explosive carbon burning at the center. However, adequate 3D computations are necessary for definite conclusions. 

\section{Conclusion}

We have carried out a detailed study of the occurrence rate of accumulation of \mc\ by CO- and ONe-white dwarf  components of semi-detached binaries. We have considered the dependence of model results on the parameter of common envelopes and  conditions for formation of the common envelopes in the intervening stages of the evolution depending on assumptions on the mixing of He-rich accreted matter. We have shown that even for the most favorable combination of the parameters accumulation of \mc\ by white dwarfs in the Galactic disk occurs at the rate that does not exceed $\simeq 2 \times 10^{-4}$\,yr$^{-1}$. Thus, in our model, this channel for formation of progenitors of potential \sna\ is not able to produce more than $\simeq 10\%$  of all \sna\ in our Galaxy. An alternative channel for formation of presupernovae -- the merger of double-degenerates is able to provide the necessary occurrence rate, but the details of the merger process itself and subsequent explosion mechanism need additional study. Among these details are, in particular, the structure and evolution of the ``core+disk'' configuration that is formed immediately after the merger, mass and angular momentum transfer in this configuration taking into account viscous processes and magnetic field, the character of burning that ensues after the merger, the role of neutrino losses and effects of rotation.

In the semi-detached systems under consideration, ``quite'' accretion-induced collapses of ONe-white dwarfs may occur with the rate up to $\sim 10^{-4}$\,yr$^{-1}$. This rate, in principle, does not contradict the limits on the rate of AIC's imposed by the considerations of the production of $r$-process elements in the Galaxy \citep{fbhc99}.

\medskip
The authors are grateful to P. Kahabka for discussion of the properties of supersoft X-ray sources. This study was partially supported by RFBR grant no. 
03-02-16254 and Federal Science and Technology Program ``Astronomy''.  

\medskip
\noindent
{\large{\bf Appendix}}

Let us make order of magnitude estimates of the time scales of several processes involved in the accretion of He and compare them with the time scale of mass exchange $t_{ex} \simeq 10^5$\,yr \citep{it85,bitzaraki02} due to  
expansion of helium remnant of the primary component of initial system. It is natural to assume that, due to shear instability, the layer of accreted matter is turbulent. Let us assume that the typical size of helium cells is equal to the local pressure scale height $H_p$ in the envelope of the star. 

The time scale of the radiative cooling of a turbulent bubble is 
\begin{equation}
t_{cool} \simeq \frac{H_p^2}{cl} \simeq \frac{H_p^2 \kappa \rho}{c},
\tag{${\text A.1}$}
\end{equation}
where $c$ is light velocity, $l$ -- mean free path of quanta,  $\kappa$ -- opacity, $\rho$ -- density. Or, in solar units,
\begin{equation}
t_{cool} \simeq 10^{-11} \kappa \rho \left( \frac{R}{\rs} \right)^4 \left( \frac{M}{\ms} \right)^2 T^2~ \text{yr},
\tag{${\text A.2}$}
\end{equation}
where $R$ and $M$ are radius and mass of the star, $T$ -- temperature. For a   2\,\ms\ star $t_{cool} \aplt 10^3$\,yr. It is evident that $t_{\textrm cool} \ll t_{ex}$. Hence, the temperature of a sinking element with the size equal to the local pressure scale height is practically equal to the temperature of the ambient matter.
 
For the equal pressure, the equal temperatures result in the difference of the density of the bubble and the ambient matter and generation of the Archimedes force. Equating the force of dynamical friction $\pi H_p^2 \rho v^2$ and the  Archimedes force $(4 \pi / 3) H_p^3 g \Delta \rho$, one may find the velocity of sinking: $ v \approx c_T \sqrt{\Delta \mu / \mu}$, where  $c_T$ is the local sound speed. Then the time of crossing the radius of the star by a turbulent element is $t_{ff} \approx (R / c_T) \sqrt{\Delta \mu / \mu}$. Since for 2\,\ms\ star $t_{ff} \simeq 2$\,hr, but the mass exchange time $t_{ex} \simeq 10^5$\,yr, one obtains that the mixing occurs practically instantaneously, if relative difference of the molecular weight of the convective cell and stellar envelope is
\begin{equation}
\frac {\Delta \mu}{\mu} = \left( \frac{t_{ff}}{t_{ex}} \right)^2 \apgt 10^{-17}.
\tag{${\text A.3}$}
\end{equation} 
Equation~(A.3) is valid always.

Further, let us estimate the rate of diffusion of helium out of convective cells. If in the process of sinking the cell does not change its identity, its typical radius varies as $r = ( \rho_0 / \rho )^{1/3} H_{p0}$, where $H_{p0} \approx 10^8$\,cm is the pressure scale height and   $\rho_0 \approx 10^{-8}$ g\,cm$^{-3}$ is the density in the photosphere. The time scale of thermal diffusion may be estimated as 
\begin{equation}
\tau_d \approx \frac{r^2}{v_{\alpha} l_{\alpha}},~~~\mathrm{where}~~~v_{\alpha} =\sqrt{\frac{3 k T}{\mu_{\alpha}}}.
\tag{${\text A.4}$}
\end{equation} 
Here $\mu_{\alpha}$ -- mass of He nuclei, $v_{\alpha}$ -- their velocity, $l_{\alpha} = 1/\sigma n_{\alpha}$ -- mean free path of the nuclei,  $n=\rho/\mu_{\alpha}$ -- concentration, $\sigma = 16 e^4 /(kT)^2$ cross-section of the interaction of nuclei, $e$ -- the charge of an electron \citep{lang74}.
Inserting $\rho_0$ and  $H_{p0}$ given above, one gets
\begin{equation}
\tau_d \approx 10^{19} \frac{\rho^{1/3}}{T^{5/2}}~~\text{yr}. 
\tag{${\text A.5}$}
\end{equation}
The time scale $\tau_d \simeq 10^2 - 10^5$\,yr  is also much shorter than the mass exchange scale for whole star, with a possible exception of the most outer layers of the envelope with relative radius of $(\Delta R / R ) \approx 0.01$.  If the turbulent fragmentation of He-enriched sinking cells occurs, $\tau_d$ has to be even shorter.

In the detailed modeling of the disk accretion of helium one has also have in mind the speed-up of stellar axial rotation and circulation of the matter caused by rotation that facilitate the mixing as well.

The circumstances considered above, lead us to conclusion that the  ``instantaneous'' mixing of whole star is quite probable and deserves further study. 

\renewcommand{\baselinestretch}{0.75}
\small

\newpage

\begin{table}
\caption[]{Parameters of the evolutionary tracks for helium-accreting star with initial mass of 2.\ms. \\ 
\small{$dM/dt$ -- accretion rate,  
$R_{max}/R_0$ and $R_{min}/R_0$ -- the ratio of maximum 
and \\ minimum values of the radius of accreting star and initial radius $R_0$, respectively.}} 
\bigskip
\begin{tabular}{c|c|c|c|c}
\hline
No. of the track & $dM/dt$,       & Mixing & $R_{max}/R_0$ & $R_{min}/R_0$ \\
 & $M_{\odot}/$yr&               &               &               \\
\hline
1 & $2 \times 10^{-5}$ & diffusion & 1.00 & 0.73 \\ 
2 & $10^{-5}$          & diffusion& 1.00 & 0.64 \\ 
3 & $10^{-6}$          & diffusion & 1.00 & 0.56 \\ 
4 & $2 \times 10^{-5}$ & instantaneous   & 2.84 & 1.00 \\ 
5 & $10^{-5}$          & instantaneous   & 1.91 & 1.00 \\ 
6 & $10^{-6}$          & instantaneous  & 1.17 & 1.00 \\ 
\hline
\end{tabular}
\label{tab:tracks}
\end{table}
\newpage
\begin{table}[ht!]
\vspace{-2.5cm}
\caption[]{Occurrence rate of potential \sna\ and of accretion-induced collapses of white dwarfs in semi-detached binaries as a function of parameters of computations. \\
{\small Observational estimate of the occurrence rate of \sna\ in the Galaxy is 
$(4\pm2)\times10^{-3}$\,yr$^{-1}$ \citep{captur01}.
For every combination of \ace\ and \at\ three upper lines give occurrence rate of possible \sna\ and AIC's in scenario I under different assumptions on the evolution of the system at the helium accretion stage: conservative mass-exchange, formation of common envelope, and threefold expansion of accretor. The fourth line gives the rate of events in systems evolving through scenario II. $N_{\rm SSS}$ is the number of supersoft X-ray sources in given model (for scenarios I and II together).
}}
\vspace{0.4cm} 
\begin{center}
{\small
\begin{tabular}{|r|c|c|c|c|l|}
\hline 
$N$ & $\alpha_{ce}$ & $\alpha_{th}$ & SN Ia & AIC & $N_{\rm SSS}$  \\
& & & $ 10^{-3}$\,yr$^{-1}$ & $10^{-3}$\,yr$^{-1}$ &   \\
\hline
1 & 1 & 1.0 & 0.065 & 0.088 & 6080   \\  
2 & 1 & 1.0 & 0.091 & 0.086 & 6050  	     \\ 
3 & 1 & 1.0 & 0.050 & 0.021 & 6010   \\ 
  &   &     & 0.001 & 0.010 &   \\    
\hline
4 & 2 & 1.0 & 0.067  & 0.002 & 3850    \\ 
5 & 2 & 1.0 & 0.098  & 0.140  & 3890	  \\ 
6 & 2 & 1.0 & 0.073  & 0.054 & 3800   \\ 
  &   &     & 0.0    & 0.003 &       \\
\hline
7 & 5 & 1.0 & 0.068  & 0.011 & 2650 \\
8 & 5 & 1.0 & 0.097  & 0.016 & 2710  \\
9 & 5 & 1.0 & 0.085  & 0.099 & 2640  \\
  &   &     & 0.0    &  0.0   &       \\ 
\hline
10 & 0.5 & 1.0 & 0.053 & 0.057 & 7660 \\
11 & 0.5 & 1.0 & 0.004 & 0.004 & 7580  \\
12 & 0.5 & 1.0 & 0.0   & 0.006  & 7580 \\
   &     &     & 0.020 &  0.032 &     \\
\hline
13 & 1 & 0.5 & 0.216 & 0.110 & 7100   \\
14 & 1 & 0.5 & 0.140 & 0.075 & 7050    \\
15 & 1 & 0.5 & 0.097 & 0.033 & 7010 \\ 
   &   &     & 0.040 & 0.024 &       \\ 
\hline
16 & 1 & 0.2 & 0.046 & 0.063 & 6750  \\
17 & 1 & 0.2 & 0.021 & 0.038 & 6700    \\
18 & 1 & 0.2 & 0.023 & 0.034 & 6700  \\
   &   &     & 0.054 & 0.033 & \\ 
\hline
19 & 2 & 0.5 & 0.210 & 0.120 & 4960   \\
20 & 2 & 0.5 & 0.220 & 0.150  & 4980     \\
21 & 2 & 0.5 & 0.150 & 0.068 & 4900  \\ 
   &   &     & 0.0   & 0.003 & \\ 
\hline
22 & 2 & 0.2 & 0.120 & 0.094 & 5240  \\
23 & 2 & 0.2 & 0.080 & 0.095 & 5210	 \\
24 & 2 & 0.2 & 0.077 & 0.066 & 5190 \\ 
   &   &     &  0.0  & 0.021 & \\ 
\hline
25 & 5 & 0.5 & 0.210 & 0.130 & 3150 \\
26 & 5 & 0.5 & 0.220 & 0.170 & 3270  \\
27 & 5 & 0.5 & 0.180 & 0.110 & 3120  \\
   &   &     &  0.0  & 0.002 & \\ 
\hline
28 & 5 & 0.2 & 0.110& 0.130& 3250 \\
29 & 5 & 0.2 & 0.110 & 0.150 & 3270 \\
30 & 5 & 0.2 & 0.091 & 0.110 & 3220 \\
   &   &     &  0.0 &  0.005 & \\ 
\hline
31 & 0.5 & 0.5 & 0.070 & 0.050 & 8630 \\
32 & 0.5 & 0.5 & 0.005 & 0.004 & 8560  \\
33 & 0.5 & 0.5 & 0.004 & 0.007 & 8560  \\
   &     &     & 0.120 & 0.046 & \\ 
\hline
34 & 0.5 & 0.2 & 0.019 & 0.033 & 7870  \\
35 & 0.5 & 0.2 & 0.0   & 0.001 & 7840 \\
36 & 0.5 & 0.2 & 0.003 & 0.007 & 7840  \\
   &     &     &  0.066 &  0.030 & \\  
\hline
\end{tabular}
}
\end{center}
\label{tab:scen}
\end{table}

\newpage
\newpage
\newpage
\newpage

\newpage

\end{document}